\RequirePackage{filecontents}

\documentclass[a4paper,fleqn,usenatbib,useAMS]{mnras}
\usepackage{graphicx}	
\usepackage{amsmath}	
\usepackage{amssymb}	
\usepackage{longtable}
\usepackage{inputenc}
\usepackage{multicol}
\usepackage{xcolor}

\usepackage{multicol}        
\usepackage{bm}		
\usepackage{pdflscape}	
\usepackage{subfig}

\usepackage[T1]{fontenc}
\usepackage{ae,aecompl}

\usepackage{multirow}

\usepackage{txfonts}
\title[The HD limit at high metallicity]{Red Supergiants in M31: \\The Humphreys-Davidson limit at high metallicity}

\author[McDonald, Davies \& Beasor]{
Sarah L. E. McDonald,$^{1}$\thanks{E-mail: S.E.McDonald@2015.ljmu.ac.uk}
Ben Davies$^{1}$ and Emma R. Beasor$^{2}$
$^{}$
$^{}$
\\
$^{1}$Astrophysics Research Institute, Liverpool John Moores University, Liverpool Science Park ic2,146 Brownlow Hill, Liverpool, L3 5RF, UK\\
$^{2}$NSF’s National Optical-Infrared Astronomy Research Laboratory, 950 N. Cherry Ave., Tucson, AZ 85719, USA\\
$^{}$
}

\date{Submitted to MNRAS, 2021 July.}

\pubyear{2021}

\begin{document}
\label{firstpage}
\pagerange{\pageref{firstpage}--\pageref{lastpage}}
\maketitle

\begin{abstract}
The empirical upper limit to Red Supergiant (RSG) luminosity, known as the Humphreys-Davidson (HD) limit, has been commonly explained as being caused by the stripping of stellar envelopes by metallicity-dependent, line-driven winds. As such, the theoretical expectation is that the HD limit should be higher at lower metallicity, where weaker mass-loss rates mean that higher initial masses are required for an envelope to be stripped. In this paper, we test this prediction by measuring the luminosity function of RSGs in M31 and comparing to those in the LMC and SMC. We find that $\log (L_{\rm max}/L_{\odot}) = 5.53 \pm 0.03$ in M31 (Z $\gtrsim$ Z$_{\odot}$), consistent with the limit found for both the LMC (Z $\sim$ 0.5 Z$_{\odot}$) and SMC (Z $\sim$ 0.25 Z$_{\odot}$), while the RSG luminosity distributions in these 3 galaxies are consistent to within 1$\sigma$. We therefore find no evidence for a metallicity dependence on both the HD limit and the RSG luminosity function, and conclude that line-driven winds on the main sequence are not the cause of the HD limit.

\end{abstract}

\begin{keywords}
stars: massive – stars: evolution – supergiants
\end{keywords}

\section{Introduction}

It is well established that there is an empirical upper limit to Red Supergiant (RSG) luminosity \citep{stothers69,sandage&tammann74}, often referred to as the `Humphreys-Davidson (HD) Limit' \citep{HD79}. The HD limit is often explained as being a manifestation of mass loss  \citep[e.g.][]{HD79} during the lifetime of the star, caused by strong stellar winds or episodic periods of mass-loss, where the fraction of mass lost from the stellar envelope is dependent on the initial mass of the star. Under this explanation, lower initial mass supergiants ($\sim8M_\odot-15M_\odot$) experience winds which are not strong enough to remove the entire hydrogen envelope on the main sequence \citep[MS,][]{maeder81,maeder&meynet03} so the star is able to evolve to the RSG phase, where it resides before dying as a core-collapse supernova. Higher initial mass stars ($\sim15M_\odot-30M_\odot$) can lose a considerable fraction of their envelope, causing the star to undergo only a brief RSG phase before evolving to a Wolf Rayet (WR) star \citep{stothers&chin79}. At even higher masses ($\gtrsim30M_\odot$), the entire envelope can be lost by the time hydrogen in the core is exhausted, preventing evolution to the cool red side of the Hertzsprung-Russell (HR) diagram. These stars instead evolve directly from the MS to a WR star, completely bypassing the RSG phase \citep{stothers&chin78}. Under this scenario, the HD limit therefore represents the luminosity which corresponds to the most massive star that may still experience a RSG phase. \newline

Massive stars lose mass both on the MS and during the RSG phase. \citet{beasor2020} show that the contribution of mass loss from cool RSG winds is extremely small, where the total mass lost is only expected to be in the range of $1-2M_{\odot}$. This means quiescent mass-loss during the RSG phase is not effective at removing a significant fraction of the Hydrogen envelope, prior to core-collapse. 

In terms of the proposed explanation of the HD limit, this then places more emphasis on mass-loss from either Luminous Blue Variable (LBV) type eruptions, discussed further in Section~\ref{massloss}, or line-driven winds during the hot MS phase \citep{cak1975,vink2001}. These line-driven winds are produced by absorption of photospheric photon momentum by UV metal lines \citep{Kudritzki2003}, and therefore it follows that there could be a metallicity dependence with radiative driven wind strength whereby decreased metallicity results in decreased wind strength \citep{Abbott1982,Kudritzki1987}. For these reasons, evolutionary models predict that lower metallicity environments should produce more luminous supergiants due to this dependency of mass loss on metallicity \citep{maeder&meynet03}. This means the HD limit should therefore also be metallicity dependent. \newline

The HD limit has been measured previously in the literature, the first being a hard upper limit of $\log(L/L_{\odot}) = 5.8 \pm 0.1$ inferred by \citet{HD79}, using an optically selected sample of cool supergiants in the Milky Way and the Large Magellanic cloud (LMC). This was later revised to $\log(L/L_{\odot}) = 5.66$ in \citet{humphreys83}. \citet[][hereafter, DCB18]{DCB18} revisited the HD limit in the Magellanic Clouds, with more complete samples and higher precision multi-wavelength photometry, finding an upper limit of {$\log(L/L_{\odot}) = 5.5$} for both the Small Magellanic Cloud (SMC) and the LMC.

To study the HD limit at higher metallicity, the most obvious environment would be the Milky Way. However, there are a number of obstacles in studying the RSG population of the Milky Way such as high foreground extinction and uncertain distances, therefore only an incomplete luminosity distribution of RSGs in the Galaxy is achievable. Although, \citet{davies&beasor20} argue that even with an effective sample size of over 100, there were still no RSGs with luminosity greater than {$\log(L/L_{\odot}) = 5.5$} in the Galaxy, and conclude the HD limit at solar metallicity is comparable to that of the SMC and LMC. However, studies of stellar populations in the plane of the Milky Way will always be subject to criticisms of completeness. Therefore, to investigate the HD limit at high metallicity, a similar galaxy-wide study to that of DCB18 is required, but in a higher metallicity galaxy.  \newline

In this paper, we complement the study of DCB18 with an investigation into the Humphreys-Davidson Limit of M31. Our close proximity to M31 \citep[0.77 Mpc,][]{karachentsev04} gives us the ability to study resolved stellar populations at a high metallicity, which is thought to be in the range of $1.05 - 1.66Z_{\odot}$ \citep{zurita&bresolin12}. The work in this current paper is distinct from other recent studies of the RSG population of M31 \citep[e.g.][]{massey&evans16,GHJ16,neugent20,massey2021} in that we focus on the high luminosity end of the RSG luminosity function and the HD limit as well as make quantitative comparisons with RSG populations in lower metallicity galaxies. \newline

\begin{figure*}
	\includegraphics[width=\textwidth]{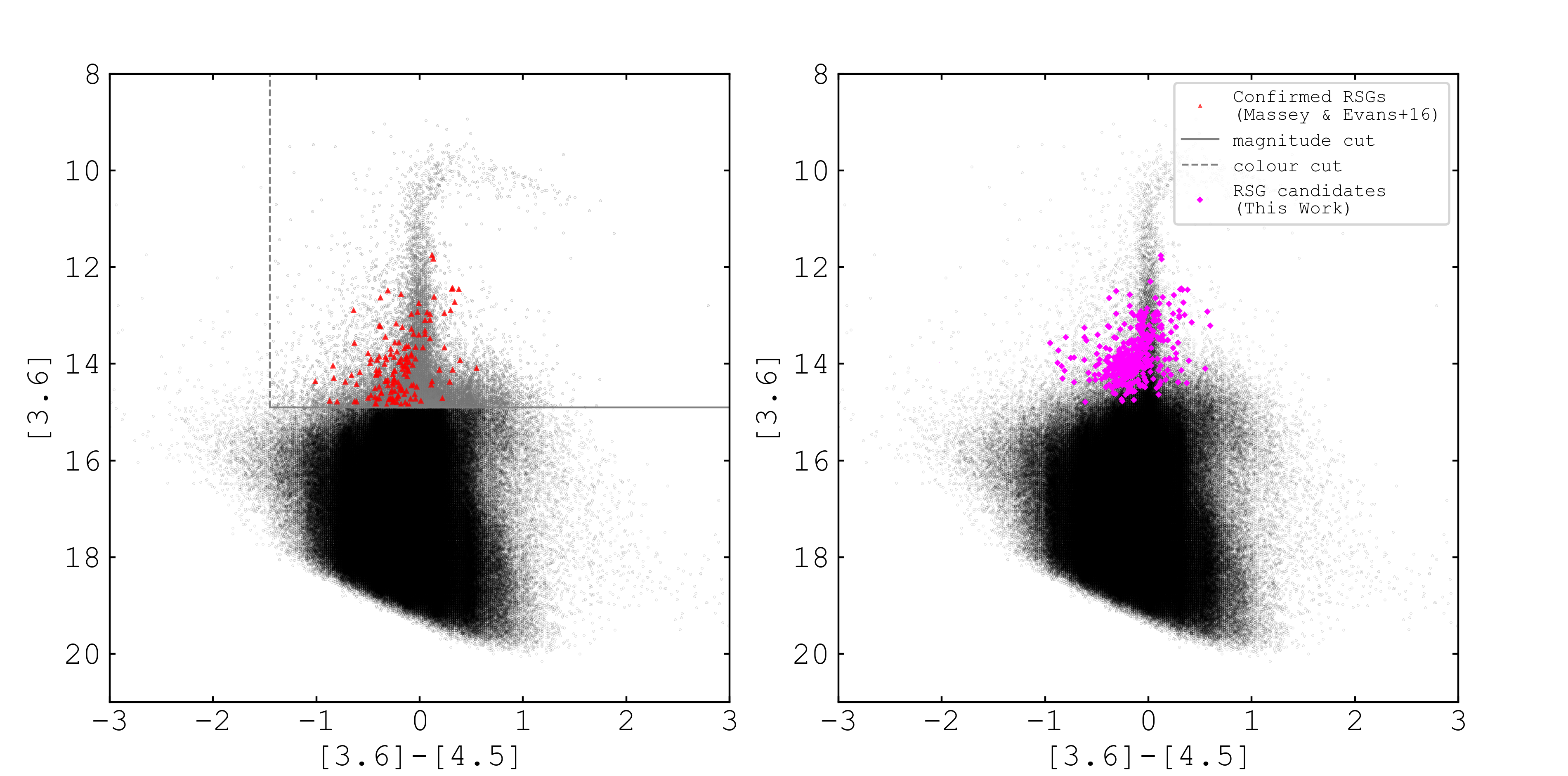}
    \caption{(a) A colour magnitude diagram, where the black points in both panels indicate all the M31 point sources detected by Spitzer IRAC/MIPS \citep{khan17}. The grey points show the sources which fit the criteria to be a likely RSG candidate based on the colour (dashed grey line) and magnitude (solid grey line) cuts applied, to find the first constraint towards establishing a sample of RSG candidates. The red triangles indicate known M31 RSGs with determined spectral classifications by \citet{massey&evans16} from which we have based our colour and magnitude cuts around. All other mid-IR cuts can be seen in Table~\ref{cuts}. \newline
    (b) The magenta points indicate all the RSG candidates (with available Spitzer mid-IR photometry) we find and use in the present work after all photometric and astrometric cuts have been applied.  }
    \label{all_CMDs}
\end{figure*}

\section{COMPILING THE SAMPLE}
\label{data}

To compile a sample of RSGs candidates in M31, we use photometry from the Spitzer mid-infrared point source survey, (IRAC/MIPS: 3.6$\mu$m, 4.5$\mu$m, 5.8$\mu$m, 8.0$\mu$m and 24$\mu$m) from \citealt{khan17}. RSGs tend to be bright in the mid-infrared as a result of their relatively low temperatures. They also experience strong stellar winds which can produce quantities of dust that can obscure stars at optical wavelengths. Therefore, any particularly dusty or `dust enshrouded' stars \citep{vanloon05} may be too faint to be detected by optical or possibly even near-IR surveys. Additionally, at these longer wavelengths there is less sensitivity to interstellar reddening. By using the Khan catalogue as a basis, we expect to have a much higher level of completeness than can be achieved from optical or near-IR surveys (more on sample completeness in Section~\ref{completeness}).
 \newline

\subsection{Method}
\label{method}
To locate our target stars, we first constructed colour-magnitude diagrams (CMDs) using the Spitzer photometry \citep{khan17}, see Figure~\ref{all_CMDs}. Next, we overplotted a sample of known RSGs from \citet{massey&evans16}, to define the location of our target stars in mid-IR colour-magnitude space. We place a colour threshold at the blue limit of known RSGs in M31, as well as a magnitude cut corresponding to log($L/L_{\odot}) \sim 4.8$ to avoid any Asymptotic Giant Branch stars (AGBs) or Red Giants contaminating our sample (\citealt{ferrari70,LambIbenHoward76,brunish86}). The colour/magnitude cuts are listed in Table~\ref{cuts}. In addition to this, we made a radius cut at 40 kpc (where the dust-free exponential disk of scale length $R_{d} = 5.3 \pm 5$ kpc,  \citealt{{Courteau2011}}), using the physical de-projected radius, assuming an inclination angle of $77.5^{\circ}$ \citep{Tempel10}.  \newline

Next we cross-matched our candidates with the RSG catalogues from \citet{massey&evans16} and \citet{GHJ16} to ensure all the brightest candidates from these optical surveys had been re-acquired through our mid-IR cuts. We found 10 objects with log($L/L_{\odot}) > 5$ from \citet{massey&evans16} and 14 from \citet{GHJ16} that do not appear in the Khan catalogue, (reasons for which are discussed in Section~\ref{determining_bolometric_luminosities}), which are then manually added in to our sample of RSG candidates. This results in a sample of 7893 RSG candidates so far. These stars are then cross-matched to the following catalogues to obtain multi-wavelength photometry and astrometry for each candidate: \newline

\begin{enumerate}
    \item Local Group Galaxy Survey (LGGS) UBVRI photometry \citet{massey06}. \newline
    \item  Gaia EDR3 photometry (BP, G and RP bands) and astrometry (proper motion and parallax) \citet{GAIADR320}.  \newline
    \item Two Micron All Sky Survey (2MASS) JHK photometry \citet{2MASS03}. 
\end{enumerate} 

After coadding the optical/near-IR photometry, we applied an extra colour criteria of Gaia Bp-Rp $>$ 1 to further screen out any objects that are too blue in colour to be RSGs.We then also use Gaia astrometry as a method of removing foreground stars. We aim for as high a completion rate as possible, so we remove any objects with a proper motion deviating more than $4\sigma$ from the motion of M31 \citep{saloman20}, to exclude foreground objects from our sample. The combination of these additional Gaia cuts remove a large number of foreground and blue objects from our sample. The total number of RSG candidates found and used in this work is 415. We discuss the completeness of our RSG sample in Section \ref{completeness}.\newline

\subsection{Correcting For Foreground Extinction}
\label{extinction}

Since we do not have associated spectroscopic information for all of these RSG candidates, we cannot correct for extinction using intrinsic colours. Furthermore, the colours of RSGs are often affected by circumstellar extinction, which unlike interstellar extinction does not reduce the observed bolometric flux (see Section~\ref{determining_bolometric_luminosities}). For these reasons, we must obtain an estimate of the foreground extinction separately. To do this we utilise an M31 extinction map \citep{dalcanton15}, surveyed by The Panchromatic Hubble Andromeda Treasury project \citep[PHAT, ][]{dalcanton12}. This provides a foreground extinction correction, $A_{v}$, for any of our RSG candidates that are situated within the north-east quadrant of M31. Each RSG candidate was then de-reddened according to the \citet{cardelli89} reddening law for the optical photometry, and \citet{Rieke&Lebofsky85} for the near-IR.\newline  

\begin{figure*}
	\includegraphics[width=1.7\columnwidth]{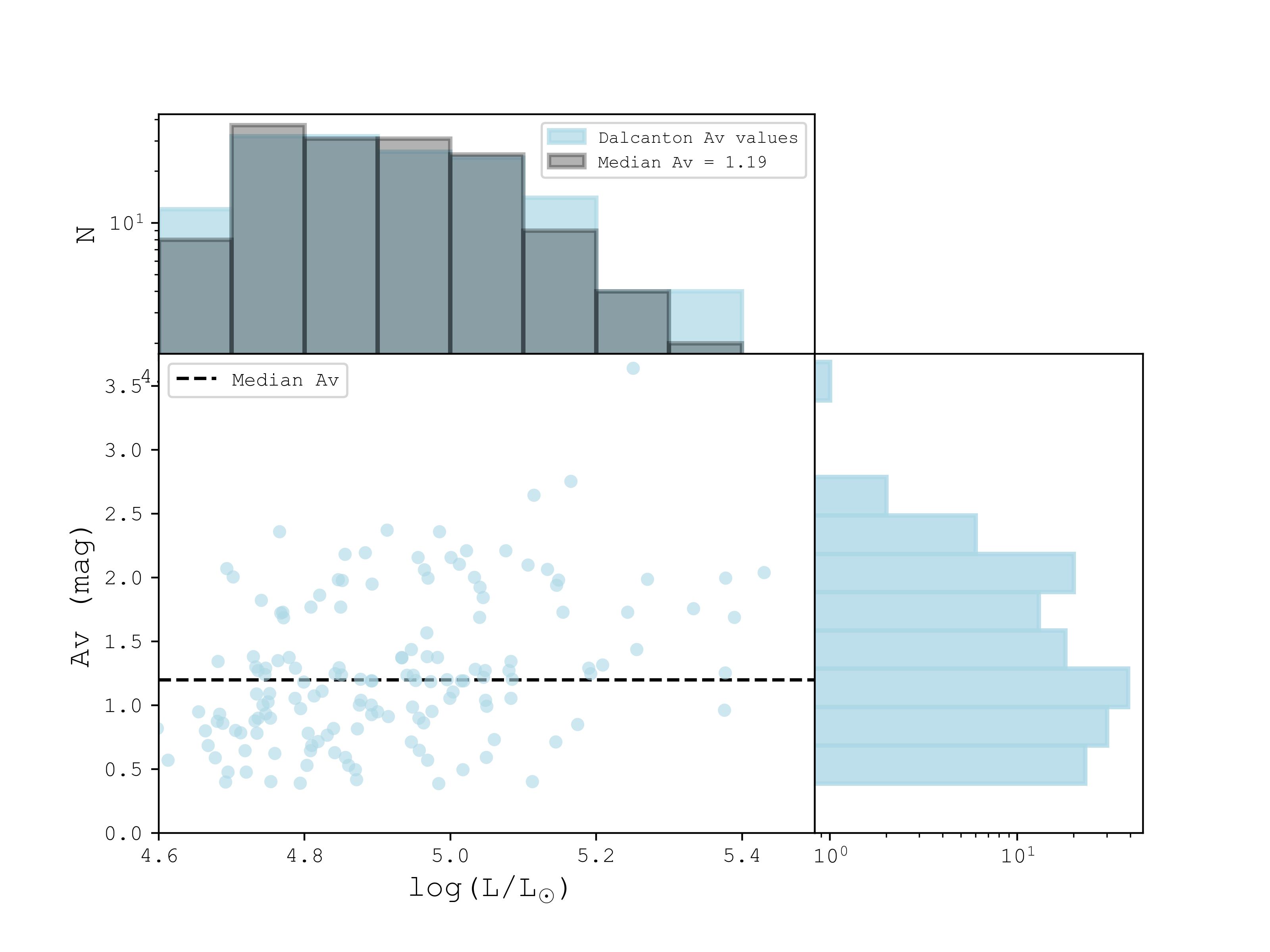}
    \caption{Top panel: A luminosity distribution of the 149 RSG candidates found in the region of M31 surveyed by HST PHAT  \citep{dalcanton12}. The luminosities in light blue are determined using $A_{v}$ taken directly from the \citet{dalcanton15} M31 extinction map. The grey distribution is the same stars but with their luminosities determined using $A_{v}=1.19\pm0.10$ which corresponds to the median of all the RSGs located within the PHAT surveyed region.
    Middle panel: Bolometric luminosity vs $A_{v}$ of 149 RSG candidates present in the M31 extinction map.\newline
    Right Panel: A distribution of the visual extinction values for each star, taken from the extinction map.}
    \label{Dalcanton_avs_lbols_comparison}
\end{figure*}

The candidates which are located outside the PHAT footprint cannot be individually extinction corrected. For these stars, we adopt the median $A_{v}$ of the 149 RSGs that are covered by PHAT. The middle panel of Figure~\ref{Dalcanton_avs_lbols_comparison} shows visual extinction $A_{v}$ of these stars as a function of the their bolometric luminosities (our method of determining bolometric luminosity is discussed in Section~\ref{determining_bolometric_luminosities}), From the median and the 68\% probability limits, we determine an average $A_{v}=1.19\pm0.10$. \newline 

To investigate whether the assumption of using a uniform $A_{v}$ for the stars not covered by PHAT introduces any systematics into our results,  we determine the bolometric luminosity of the 149 candidates using both their individual $A_{v}$ from the extinction map and the median $A_{v}=1.19$. The top panel of Figure~\ref{Dalcanton_avs_lbols_comparison} shows that the number of objects in each bin of the luminosity function when using both the average uniform $A_{v}$ and the individual PHAT extinction corrections. Though the exact number of objects in each bin is different, the two are consistent to within the Poisson errors. Furthermore, $L_{\rm max}$ is the same whichever extinction correction method is used. Therefore, we conclude that the assumption of a uniform $A_{v}$ results in a luminosity distribution and $L_{\rm max}$ which are stable to within the error margin.   \newline

\begin{table*}
\begin{tabular}{l|l}
\hline
Spitzer magnitudes (IRAC/MIPS) & Magnitude cut (mags) \\
\hline
IRAC1 (3.6$\mu$m)              & 14.9                 \\
IRAC2 (4.5$\mu$m)              & 15.0                 \\
IRAC3 (5.8$\mu$m)              & 14.8                 \\
IRAC4 (8.0$\mu$m)              & 14.8                 \\
MIPS1 (24$\mu$m)               & 12.8                 \\
\hline
Spitzer colours (IRAC/MIPS)    & Colour cut (mags)    \\
\hline
{[}3.6{]} - {[}4.5{]}            & -1.45                \\
{[}3.6{]} - {[}5.8{]}            & -1.3                 \\
{[}3.6{]} - {[}8.0{]}            & -1.0                 \\
{[}5.8{]} - {[}8.0{]}            & -0.6                 \\
{[}3.6{]} - {[}24{]}             & 0.0                  \\
{[}8.0{]} - {[}24{]}             & 0.0                  \\
{[}4.5{]} - {[}24{]}             & 0.4                 \\
\hline
\end{tabular}
\caption{The Spitzer colour and magnitude cuts that were applied to locate our target stars. The cuts are based on the colours and magnitudes of known confirmed RSGs from \citet{massey&evans16}, indicated by the red triangles. All the Spitzer point source detections for M31 are shown in black.}
\label{cuts}
\end{table*}

\section{Luminosity Distributions and $L_{\rm max}$}
\label{luminosities}
\begin{figure*}
	\includegraphics[width=0.8\textwidth]{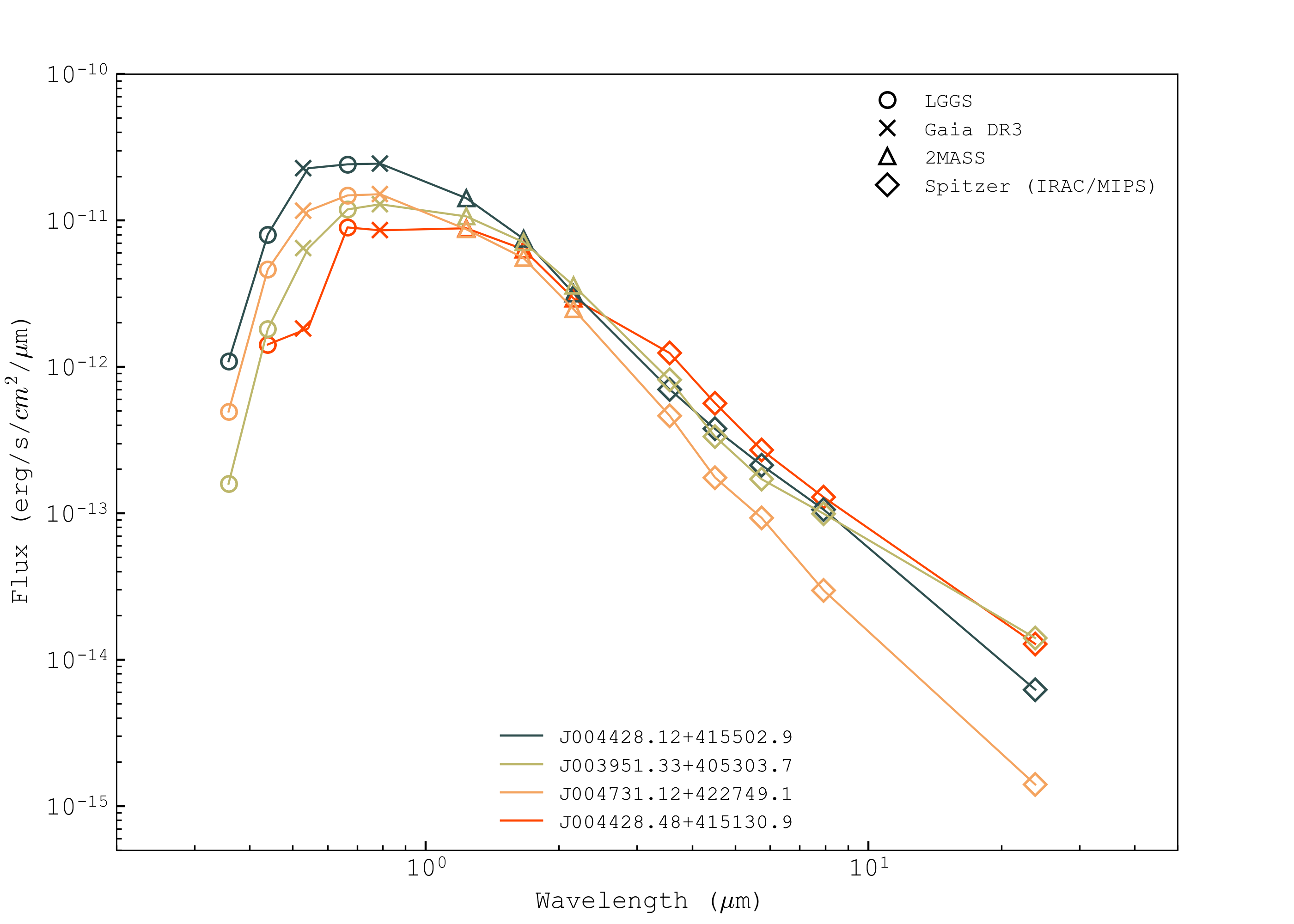}
    \caption{Spectral energy distributions of the most luminous Red Supergiant (RSG) candidates. These have log$(L/L_{\odot}) > 5.4$ with complete de-reddened photometry ranging from the optical through to the mid-infrared. The symbols in the upper legend indicate the catalogue source of the photometry and the lower legend provides the LGGS star name for each candidate. Each of these luminous RSGs are discussed in more detail in Section~\ref{mostluminous}. }
    \label{sed}
\end{figure*}

\subsection{Determining Bolometric Luminosities}
\label{determining_bolometric_luminosities}
We converted the de-reddened photometry into fluxes using Vega calibrated zero point fluxes for each filter from the SVO Filter Profile Service \citep{SVO}. Using these fluxes, we plot spectral energy distributions (SEDs) for each RSG candidate and integrate under the SED to determine bolometric luminosity, using IDL routine \texttt{int\textunderscore tabulated}, adopting an M31 distance modulus of 24.4 \citep{karachentsev04}. In doing so, we make the same assumption as DCB18 that any flux lost to absorption by circumstellar material is re-radiated at longer wavelengths, and so by integrating under the SED from the optical to the mid-IR we obtain all the star's flux. Figure~\ref{sed} shows the SEDs of the most luminous candidates with complete photometry from optical to the mid-IR. In Section~\ref{mostluminous}, we discuss in more detail the brightest RSG candidates as well as any bright objects which were rejected from our sample. \newline

\begin{figure*}
	\includegraphics[width=\textwidth]{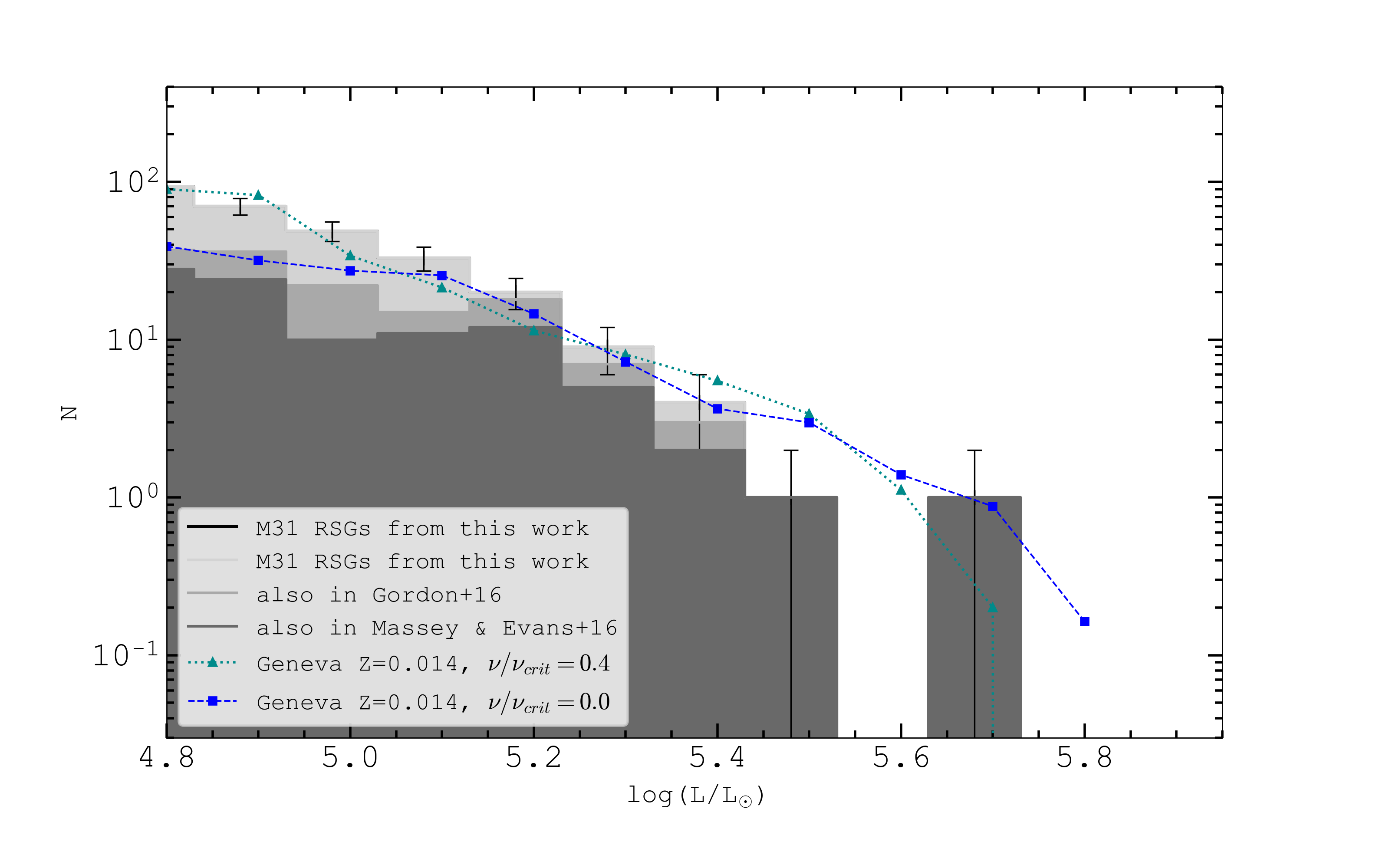}
    \caption{The Red Supergiant luminosity distribution for M31. The observed luminosity distribution from this work is shown in light grey, with the two darker grey distributions showing the number of RSG candidates we use in this study that are also found in previous M31 RSG studies. Over-plotted are the rotating ($\nu/\nu_{crit}=0.0$) and non-rotating ($\nu/\nu_{crit}=0.4$) model predicted distributions from the GENEVA models at solar metallicity (Z=0.014) from \citet{ekstrom12}.  N.b. The brightest star at log$(L/L_{\odot}) = 5.71$ cannot be definitively ruled out, but is a borderline M31 candidate due to its proper motion. This is discussed further in Section~\ref{mostluminous}.}
    \label{Ldist}
\end{figure*}

\begin{figure}
	\makebox[\columnwidth][c]{\includegraphics[width=1.25\columnwidth]{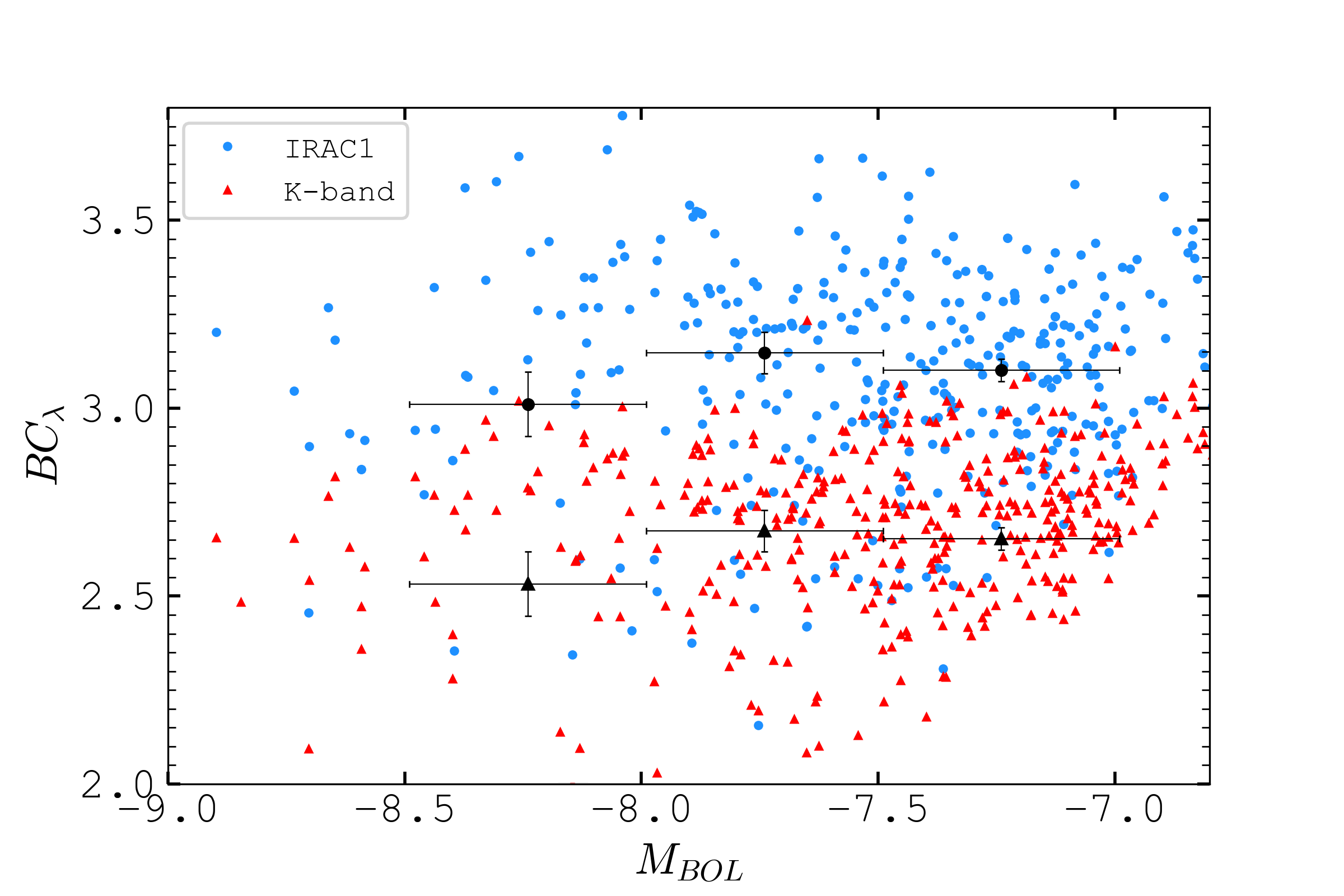}}
    \caption{The offset $M_{bol}$ - $M_{\lambda}$ is used to estimate the bolometric correction for each RSG candidate, in both the K-band seen in red and IRAC1 $3.6\mu$m band, in blue.}
    \label{bolcorrection}
\end{figure}

A few of the objects in our sample have incomplete photometric coverage, often due to them being undetected at longer wavelengths. These objects were identified when comparing the sample of stars found in the present work with previously compiled M31 RSG catalogues  \citep{massey&evans16,GHJ16}), where we found 24 stars with log$(L/L_{\odot}) > 5$ were missed from our study. Spitzer/IRAC and MIPS images \footnote{Spitzer images were taken from the NASA/IPAC Infrared Science Archive} of these objects show they appear to be spatially extended in the mid-IR, and as a result are absent from the \citet{khan17} point source catalogue. To estimate luminosities of these objects, we employ an alternative method of using a K-band bolometric correction ($BC_{K}$) which we describe below. \newline

\subsection{Bolometric corrections}
In this section, we use the RSGs with complete photometric coverage to determine bolometric corrections (BCs) appropriate for M31. We use K-band photometry since a $BC_{K}$ at this wavelength does not appear to be sensitive to spectral type (DCB18). Furthermore, the extinction at this wavelength is only around 1/10th of that in the V-band. The bolometric correction is then used to estimate luminosities for the stars with incomplete SED coverage. \newline

We individually de-redden the near-IR photometry, prior to converting to bolometric luminosity by employing either (a) a uniform $A_{K}=0.13\pm0.02$ found from the median $A_{v} =1.19\pm0.1$ and the relation A$_{K} = 0.11A_{v}$ from \citet{Rieke&Lebofsky85}; or (b) the individual A$_{v}$ if situated in the extinction map and finding A$_{K}$ using the same $A_{V}/A_{K}$ relation. We then calculate BC by finding $M_{BOL}-M_{\lambda}$ for each of our RSG candidates, see Figure~\ref{bolcorrection}. We find the median $BC_{K} = 2.71\pm0.12$ as well as $BC_{IRAC1}=3.18\pm0.15$, where the uncertainty is the standard deviation. We also plot a binned average of $BC_\lambda$ with $M_{BOL}$ to show that within the uncertainties there is no systematic trend with brightness. \newline

The $BC_{K}$ we find for M31 is consistent with those found in  previous studies for other local group galaxies. There is a good agreement with the median BC across spectral classes K and later derived for the LMC, with a median BC of 2.81$\pm$0.08 and SMC with 2.60$\pm$0.09 both from DCB18 as well as 2.81$\pm$0.10 for the Milky Way from \citet{db18}. \newline

\begin{table*}
\renewcommand{\arraystretch}{1}
\centering
\resizebox{0.7\textwidth}{!}{%
\begin{tabular}{llllll}
\hline 
LGGS Name  &  RA DEC (J2000)  & log$(L/L_{\odot})$ &  Classification       \\
\hline
J004520.67+414717.3  & 00 45 20.66	+41 47 17.1 &   $5.75\pm0.11$* &    RSG (M1I)$\dagger$ \\

J004428.12+415502.9 &  00 44 28.11	+41 55 02.7 & $5.53\pm0.03$ &  RSG (K2I)           \\

J004539.99+415404.1 &  00 45 39.98	+41 54 03.9 & $5.49\pm0.09$*  & RSG (M3I)            \\

J003951.33+405303.7 &  00 39 51.32	+40 53 03.6 & $5.46\pm0.02$   & `Possible' RSG       \\

J004731.12+422749.1 & 00 47 31.04	+42 27 48.2 & $5.44\pm0.04$ & `Possible' RSG   \\

J004428.48+415130.9 &  00 44 28.47	+41 51 30.7 & $5.43\pm0.02$   & RSG  (M1I)           \\

\hline
\end{tabular}
} 
\caption{The name, position and bolometric luminosity of the RSG candidates with log$(L/L_{\odot}) > 5.4$ found in this study. We also provide the SIMBAD object classification of each candidate, assigned by \citet{massey09}/\citet{massey&evans16}. Full analysis of these objects and their luminosities are described in Section~\ref{mostluminous}. *The uncertainty of these luminosities is dominated by the error on the $BC_{K}$, discussed further in Section~\ref{determining_bolometric_luminosities}.$\dagger$This is our borderline candidate which has been previously classified as an M1I supergiants but our caveats for this objects are discussed in Section~\ref{mostluminous}.}
\label{brightest_stars}
\end{table*}

\subsection{The Most Luminous RSG candidates in M31}
\label{mostluminous}

The most important candidates for our investigation into the HD limit are those occupying the high end of the luminosity function. The photometric and astrometric constraints implemented, previously discussed in Section~\ref{data}, ensure the stars in our sample have the appropriate colours, magnitudes and proper motion consistent with being RSG candidates in M31. However, only approximately $25\%$ of the sample has spectroscopic confirmation. This means that there may be some contamination. Therefore, a further verification step we applied was to inspect high spatial resolution archival images such that all objects at the bright end of the luminosity function (log$(L/L_{\odot}) > 5.3$) are consistent with being single sources. \newline

The observational luminosity function of M31 RSGs is shown in Figure~\ref{Ldist} by the light grey distribution. It shows the number of RSG candidates per log luminosity bin for M31, found in the present work. The two darker grey distributions show the number of RSG candidates we use in this study which are also found in previous M31 RSG studies. For the brightest RSGs, their luminosities and spectral classifications can be seen in Table~\ref{brightest_stars}. Below are the most luminous candidates discussed in more detail. \newline

\begin{enumerate}

   \item \textbf{J004520.67+414717.3}:  This object has previously been assigned a spectral classification of M1I with a luminosity of $\log$(L/L$_{\odot}) = 5.81$ by \citet{massey&evans16} and $\log$(L/L$_{\odot}) = 5.94$ by \citet{GHJ16}. In the present work, we determine a luminosity of $\log$(L/L$_{\odot}) = 5.75\pm0.11$. This makes this object the brightest RSG candidate we find in M31. However, there are some caveats to the significance of this high luminosity object in regards to $L_{\rm max}$. Firstly, although this object has optical colours consistent with RSGs (B--V = 2.68 and V--R = 1.55), the source appears to be sat within spatially extended infrared emission, meaning that it does not appear in the Khan point-source catalogue. As a result its luminosity is determined using a $BC_{K}$ where the large uncertainty is dominated by that on the bolometric correction. Further, this candidate has a proper motion which deviates from the M31 proper motion at the 2$\sigma$ level. This raises the possibility that it is a foreground object, which is also suggested in \citet{massey&evans16}, as they find the radial velocity of this object overlaps foreground star velocities. This casts further uncertainty on its luminosity as the M31 distance assumed in the luminosity calculation is no longer appropriate if not an M31 member. Since we cannot definitively rule out this object since it has a RSG classification, it remains in our sample. However, we will treat this object with caution in regards to the HD limit. \newline

    \item \textbf{J004428.12+415502.9}: This candidate has been previously classified as a K2I RSG from \citet{massey&evans16} with a luminosity of $\log(L/L_{\odot}) = 5.64$, as well as a luminosity of $\log(L/L_{\odot}) = 5.89$ from \citet{GHJ16}. It has also been described as a Long Period Variable candidate in \citet{Sorisam18} in their study of RSG variability in M31. We initially found a luminosity of {$\log(L/L_{\odot}) = 5.63$}, but closer inspection of SDSS images show it to be two blended stars of similar colour. The brighter of the two stars has astrometry consistent with M31, but the fainter has a high proper motion and is therefore likely to be a foreground object. From the ratio of the two stars' fluxes, we estimate that the M31 star has an apparent brightness 0.1dex greater than the foreground star. This leads to a revised brightness for the RSG of {$\log(L/L_{\odot}) = 5.53\pm0.03$}. \newline   
    
      \item \textbf{J004539.99+415404.1}: This star is classified as a M3I RSG with a luminosity of $\log(L/L_{\odot}) = 5.81$ in \citet{massey&evans16} and $\log(L/L_{\odot}) = 6.09$ in \citet{GHJ16}. In the present work, we initially calculated a luminosity of $\log(L/L_{\odot}) = 5.81$ from its SED, but in HST images and in Gaia DR3 we find that the object resolves into two sources. One source has no Gaia astrometry but the other has a large detectable proper motion in Gaia DR3, indicating foreground membership.The two sources also have comparable brightnesses and colours at Gaia Bp and Rp wavelengths (the RSG candidate: Bp-Rp = 2.389475 and the nearby red object: Bp-Rp = 2.389486. We have full SED coverage for this object, but the derived luminosity will consequently contain flux contributing from both sources in the near and mid-IR, which results in an over-estimation of the objects luminosity. Under the assumption that the star with no astrometry is an M31 member, and that the stars are of comparable apparent brightness at all wavelengths, we use the 2MASS K-band photometry (which detects these objects as only one source) and allocate a K-band flux to the RSG, that is half of the total K-band flux. We then use a K-band BC to determine its luminosity, which we find to be $\log(L/L_{\odot}) = 5.49\pm0.09$. \newline

    \item \textbf{J003951.33+405303.7}: This candidate has been previously identified as a `possible RSG' in \citet{massey09}, but has not been spectroscopically confirmed. This object has optical colours consistent with RSGs and in SDSS images appears as a single object for which we find a luminosity of {$\log(L/L_{\odot}) = 5.46\pm0.02$}. This object also passed our proper motion constraint of deviating less than 4sigma from the proper motion of M31, consistent with M31 membership. We find no reason to exclude this object based on its high resolution images, therefore it remains in our sample. \newline
    
    \item \textbf{J004731.12+422749.1}: This object is a `possible RSG', according to \citet{GHJ16} with a luminosity of $\log(L/L_{\odot}) = 5.53$ but has not been spectroscopically confirmed. This object passed our proper motion cuts and is therefore presumed to be an M31 member. In the present work, we determine a luminosity of $\log(L/L_{\odot}) = 5.44\pm0.04$ and find no reason to reject this object and so it remains in our sample. \newline

    \item \textbf{J004428.48+415130.9}: This is another confirmed RSG with a spectral type of M1I and a previously determined luminosity of $\log(L/L_{\odot}) = 5.60$ by \citet{massey&evans16} and $\log(L/L_{\odot}) = 5.64$ by \citet{GHJ16}. The Gaia proper motions of this object are consistent with the proper motion of M31, so we presume that this object is an M31 member. Lastly, this star appears to be a single object in HST images for which we find a luminosity of $\log(L/L_{\odot}) = 5.43\pm0.02$ from its SED.

\end{enumerate}

\subsubsection{Stars rejected from this work}
\label{rejected_this_work}
The following objects are those that met both our colour and magnitude criteria and have $\log(L/L_{\odot}) > 5.3$, but were rejected after inspecting their high resolution images. The reasons for rejection are described below: \newline

\begin{enumerate}
   \item \textbf{J004257.58+411740.1}: Our initial estimate of this star's luminosity was $\log(L/L_{\odot}) = 5.81$. However, despite having both mid-IR and optical colours consistent with RSGs, this object is located within the bulge of M31 where there is little to no star formation occurring making it unlikely to be a massive star. Also, the object appears spatially extended in HST B-band images, consistent with the object being a globular cluster, which is also suggested by \citet{wirth85}. Therefore, we reject this object from our sample. \newline
   
 \item  \textbf{J004336.68+410811.8}: This object appears in the \citet{GHJ16} sample, estimated to have a luminosity of $\log(L/L_{\odot})=5.86$. It is also mentioned in the \citet{massey&evans16} study as a possible RSG but has no derived luminosity due to the object having no K-band photometry. However, the object is resolved in HST U-band imaging, showing that it is instead a star cluster. The object was rejected from our sample. 
\end{enumerate}

\subsection{Sample completeness}
\label{completeness}
Inferring an upper luminosity limit of cool supergiants is difficult due to the steep power law present in the RSG luminosity function, as a result of both the initial mass function (IMF) and the short lifetimes of massive stars. This means low number statistics have a strong influence on our results, as $L_{\rm max}$ is extremely sensitive to sample size (discussed in more detail in Section~\ref{Comparisons_to_theoretical_predictions_of_lmax}). Therefore we aim to ensure sample completeness for all RSGs with $\log(L/L_{\odot}) > 5$, since we are focused on the high end of the RSG luminosity function and the HD limit. Below this luminosity, we are at more risk of including contaminating objects. To aim for completeness, as mentioned in Section~\ref{method}, we cross-checked our sample with other M31 RSG catalogues which instead optically select their RSGs, to check all previously identified RSGs were acquired through our mid-IR cuts. There were, however, a small number of objects that were missing from the Khan catalogue, as previously discussed in Section \ref{determining_bolometric_luminosities} which are sat in spatially extended mid-IR emission, meaning that they are not point-sources in the mid-IR. Therefore, the only RSGs that could be missed by our sample selection are those which are faint in the optical (e.g. due to circumstellar dust) but also spatially extended in the mid-IR due to confusion with other nearby sources, and hence missing from the point-source catalogue. Any objects absent from the mid-IR point source catalogue, but were bright in optical wavelengths were manually added to our sample. All RSG candidates found in the present work which were \textit{also} found in previous studies can be seen in Figure~\ref{Ldist}. The total number of RSG candidates we found in this study is 415, although for the statistical analysis carried out in the present work (See Section~\ref{Comparison_to_lower_metallicities} onward) we take our sample size to be the 117, which is the number of RSGs with $\log(L/L_{\odot}) > 5$. \newline


In \citet{massey&evans16}, they have a sample of 251 M31 RSGs with assigned spectral classifications, where 50 of these have a luminosity greater than $\log(L/L_{\odot}) > 5$. From their sample we have re-acquired all 50 of those with $\log(L/L_{\odot}) > 5$ in our sample. \newline

The total number of RSGs with $\log(L/L_{\odot}) > 5$ in \citet[][hereafter,GHJ16,]{GHJ16} is 139. We re-acquired 128 of these either with our cuts or were manually added to our sample if not present in the \citet{khan17} catalogue. The remaining 11 objects were inspected in HST and SDSS imaging, and in each case, we found justification for rejecting them from our sample. The reasons for rejection in each of these 11 individual cases are discussed in Section~\ref{rejected_from_ghj16_me16}. \newline

This means our sample contains all the known RSGs in M31 with $\log(L/L_{\odot}) > 5$ from previous work as well as 48 candidates which we found through our own colour/magnitude criteria\footnote{It should be noted that when we calculated the luminosities using our SED method of all the stars from previous work, some had revised luminosities meaning that they no longer had luminosities greater than $\log(L/L_{\odot}) > 5$, hence our sample size greater than $\log(L/L_{\odot}) > 5$ is smaller than in \citet{GHJ16}.} \newline

\subsubsection{Rejected stars from \citet{massey&evans16} and \citet{GHJ16}}
\label{rejected_from_ghj16_me16}
Below are the objects from previous M31 RSG catalogues which we have rejected from our study: \newline

\begin{enumerate}
\item \textbf{J004105.97+403407.9, J004431.71+415629.1, J003942.43+403203.5 and J003811.56+402358.2}: These objects from GHJ16 have Gaia EDR3 proper motions which indicate they are foreground objects, deviating from M31's proper motion by $\sim3-4\sigma$. \newline

\item \textbf{J004303.21+410433.8 and J004052.19+403116.6}:
These two objects are present in the GHJ16 sample but have assigned spectral types of B0.5I and B8, respectively, found in \citet{revised_ubvri_2016}. \newline 

\item \textbf{J004416.28+412106.6 
and J004259.31+410629.1}: For the object J004416.28+412106.6, although described as an RSG candidate in GHJ16, both \citet{revised_ubvri_2016} and \citet{HIIregion11} classify this as an HII region. It also has a low B--V colour of 0.22 which corresponds to a spectral classification much earlier than K or M. Similarly for J004259.31+410629.1, this object has a low B--V colour of 0.70 which again suggests an early spectral type. Gaia EDR3 also shows J004259.31+410629.1 to have a huge proper motion ($19.8$ mas yr$^{-1}$), suggesting that it is not an M31 object. Additionally, \citet{Soraisam2020} discuss that not only is this object located within an HII region, it also shows characteristics of being W-Ursae-Majoris contact binary, with J004259.31+410629.1 being foreground. \newline

\item \textbf{J003948.45+403131.5}: This object has a Gaia Bp-Rp colour of 0.73 which means it does not meet our red criteria. It is absent from the \citet{khan17} mid-IR catalogue for M31 due to being spatially extended and has crowded LGGS photometry, which Gaia EDR3 is unable to resolve. It is described as a young cluster in \citet{Caldwell09,Kang2012} and is therefore rejected from our sample. \newline

\item \textbf{J004331.04+411815.9} and \textbf{J004336.68+410811.8}: These two objects are both located in the halo of M31 and appear to be spatially extended in HST PHAT images, which suggests that they are possibly globular clusters. \newline

\end{enumerate}

\section{Discussion}
\label{discussion}
\subsection{Comparison with previous work}

\begin{figure}
	\includegraphics[width=1.2\columnwidth]{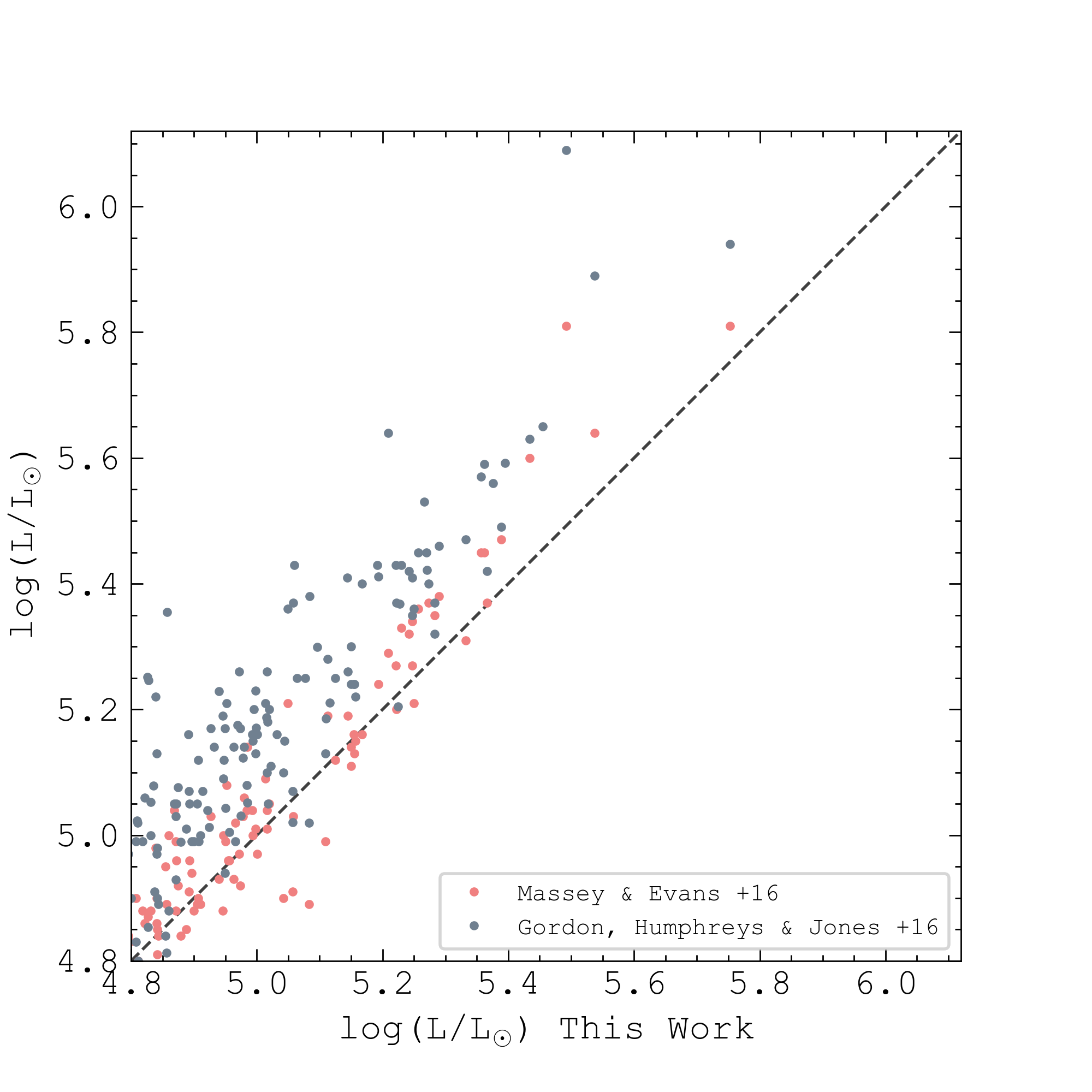}
    \caption{Comparison of bolometric luminosities found in the present work and previous studies of M31 RSGs. The grey points denote the luminosities from this work compared with \citet{GHJ16} and pink points shows the comparisons of this work with the luminosities from \citet{massey&evans16}. The black dashed line indicates the 1:1 line. }
    \label{lum_compare}
\end{figure}

\begin{figure*}
	\includegraphics[width=\textwidth]{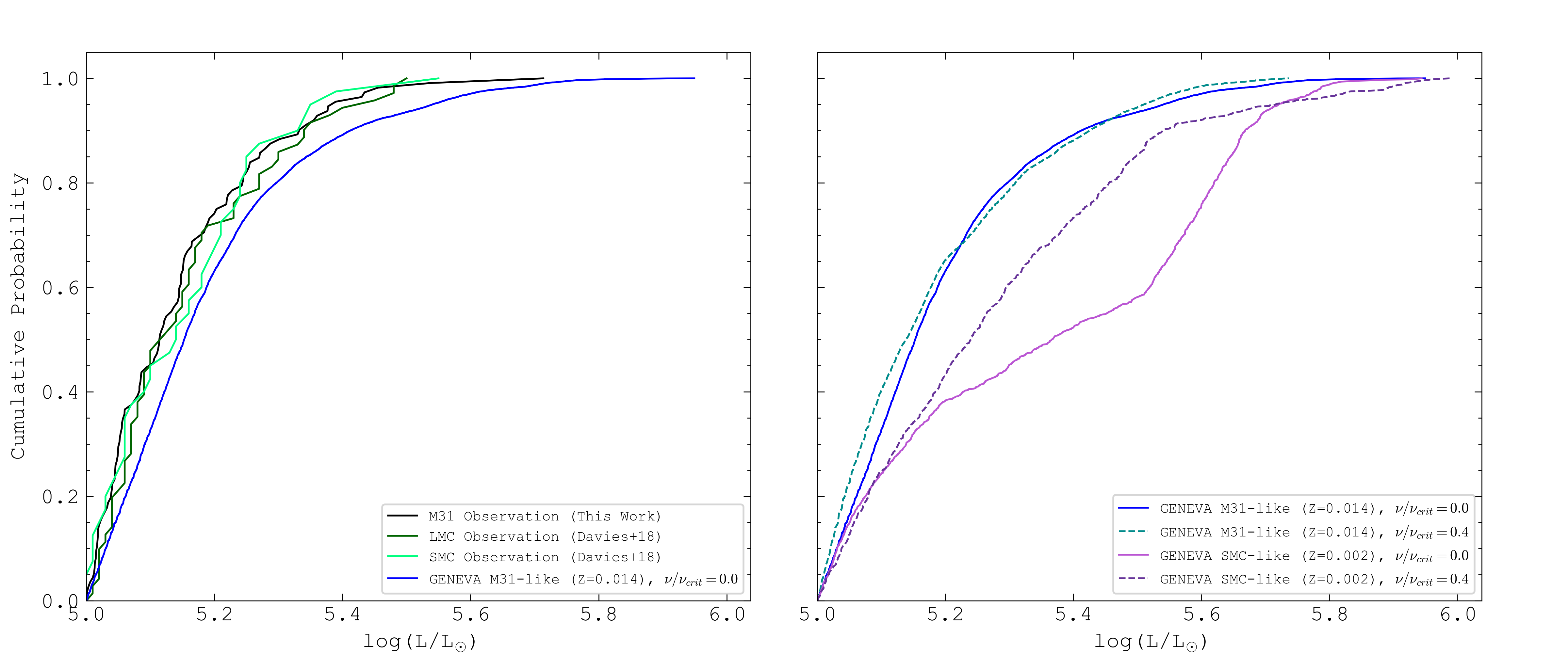}
    \caption{Left Panel: The cumulative luminosity distribution of all the Red Supergiants with an observational luminosity $\log(L/L_{\odot})>5$ in M31 from this work, as well as for the Large and Small Magellanic Clouds from \citet{DCB18}.  \newline
    Right Panel: Cumulative luminosity distribution of the cool supergiants with a luminosity $\log(L/L_{\odot})>5$ from the model luminosity functions predicted by GENEVA, at both solar \citep[from][]{ekstrom12} and SMC-like metallicities \citep[from][]{Georgy2013} for both the rotating and non-rotating models. We include the `M31-like' non-rotating model predicted distribution in the left panel for comparison.}
    \label{2_panel_ks_test}
\end{figure*}

Our results show that the luminosities determined in the present work are on average lower compared to those found for the same stars in previous work, especially those at the high end of the luminosity function. In particular, for the stars J004539.99+415404.1,  J004520.67+414717.3 and J004428.12+415502.9 \citet{massey&evans16} find {$\log(L/L_{\odot})$} = 5.81, 5.81 and 5.64, respectively. These are 0.32 and 0.06 and 0.11dex brighter than found in the present work. The same is seen when compared to \citet{GHJ16}, where they find {$\log(L/L_{\odot})$} of 6.09, 5.94 and 5.89, which is 0.60, 0.19 and 0.36dex brighter than this work. This is shown in Figure~\ref{lum_compare}, where there is both a systematic offset between the luminosity samples as well as object to object differences. Below we describe the differences between these studies in more detail. \newline

\subsubsection{Comparison with \citet{massey&evans16} and \citet{neugent20}}
We find our luminosities to be broadly consistent with those found from \citet[][hereafter, ME16]{massey&evans16}, however there is a disagreement when it comes to the higher luminosity RSGs. We include comparisons with more recent work by \citet{neugent20}, who also measure the RSG luminosity function in M31 and adopt a few of the same techniques as ME16 such as, extinction correction method and the use of a $BC_{K}$ to determine bolometric luminosity. Here we discuss the possible reasons for differences in luminosity for these objects. \newline

Firstly, to correct for foreground visual extinction $A_{v}$ in ME16, they adopt a uniform value of $A_{V}=1$ derived from their spectral fits to optical spectrophotometry of each RSG in their sample. Later work by \citet{neugent20}, use the same approach but also introduce a brightness-dependent extinction component which causes the brightest RSGs to have extinctions proportional to their K-band brightnesses. This then has the effect of systematically shifting the brighter RSGs to higher luminosities and warmer temperatures, which leads to a higher $L_{\rm max}$ of $\log(L/L_{\odot}) \approx 5.7$, compared to when adopting a uniform $A_{v}=1$, which results in a reduced $L_{\rm max}$ of approximately $\log(L/L_{\odot})\approx 5.5$. Though \citet{massey2021} comment that using this added extinction component leads to `much better agreement with the evolutionary tracks' and than would have occurred by adopting a uniform $A_{v}$,  our goal in the present work is to test these same evolutionary models. Therefore, for us to use these models to inform our choice of extinction correction would be circular logic on our part. Instead, we employ an independent method to estimate each star's extinction, specifically, through the use of an M31 extinction map \citep{dalcanton15} and adopting the median $A_{v}=1.19\pm0.10$ for those not covered by the map. Therefore, the extinctions we assign to the brightest objects are inevitably lower than those adopted by \citet{neugent20}. \newline
 
To obtain bolometric fluxes, ME16 employ the $T_{\rm eff}-BC_{K}$ relation, derived from fitting MARCS model atmospheres to optical spectra from \citet{massey09}. However, it is well known that these model atmospheres perform poorly at optical wavelengths, leading to systematic errors in $T_{\rm eff}$ and in $BC_{K}$ \citep[see][]{Davies13}. Our method of estimating $L_{\rm bol}$ from integrating the SED is free of any such model dependencies.  \newline 
 
Another factor which directly affects their luminosities is HST PHAT and Gaia EDR3 showing some of their most luminous RSGs resolving into multiple sources. In the present work, we have flagged that both J004539.99+415404.1 and J004428.12+415502.9 resolve into two objects, both with one source having a proper motion inconsistent with M31 and the other being a likely M31 member. When we account for the luminosity of the blended stars in these cases, it results in a downward revision of the our original SED derived $L_{\rm bol}$.

\subsubsection{Comparison with \citet{GHJ16}}
The bolometric luminosities calculated for the RSG candidates in \citet[][hereafter, GHJ16]{GHJ16} are on average 0.16dex higher than those found for the same stars in the present work. Figure~\ref{lum_compare} shows the systematic offset in luminosity between the two studies. In GHJ16, they adopt the same approach of integrating SEDs to obtain $L_{\rm bol}$, although they only integrate from the optical to the near-IR K-band, unless there is evidence for circumstellar dust where they then integrate out to the $22\mu$m WISE band. However, the WISE mid-IR photometry has limited angular resolution, which can result in incorrect cross-identification of objects in different catalogues, as highlighted by GHJ16 themselves. An informative example of this is the source J004539.99+415404.1, which appears in the GHJ16 catalogue as having $\log(L/L_{\odot})=6.09$. In their analysis, GHJ16 employ photometry from the ALLWISE catalogue across all four bands. However, inspection of the WISE images at $12\mu$m and $22\mu$m reveal that there is no point source at this position. Instead, at these wavelengths we see only the bright background emission of the underlying spiral arm which is incorrectly attributed to the RSG in the ALLWISE point-source catalogue. This phenomenon is responsible for GHJ16 overestimating the luminosities of many objects in their sample. \newline

A second difference between the present work and GHJ16 is how extinction is accounted for. GHJ16 explore two separate methods: firstly, they estimate $A_{V}$ from colours of nearby O and B type stars; and secondly, they derive $A_{V}$ from the relation between neutral hydrogen column density and the colour excess E(B--V) along the line of sight to each RSG candidate. However, since a large fraction of their RSGs have no nearby OB stars, not all of their RSGs have $A_{V}$ estimates from both methods, where $\sim 67\%$ of their stars have HI-based $A_{V}$ estimates only. In the circumstances where there is an extinction measurement available via both methods, the OB star method is favoured. However, this method often yields a much larger $A_{V}$ compared to their alternate method. One example being, the RSG candidate J004304.62+410348.4, for which they find {$\log(L/L_{\odot})$} = 5.40 with $A_{V}$ = 2.1 from their OB colour method which they adopt, but they also find and $A_{V}$ = 1.3 from the neutral hydrogen column density method. Using these higher extinction values contribute to the higher luminosities for these stars.
\newline

\begin{figure*}
	\centerline{\includegraphics[width=1.1\textwidth]{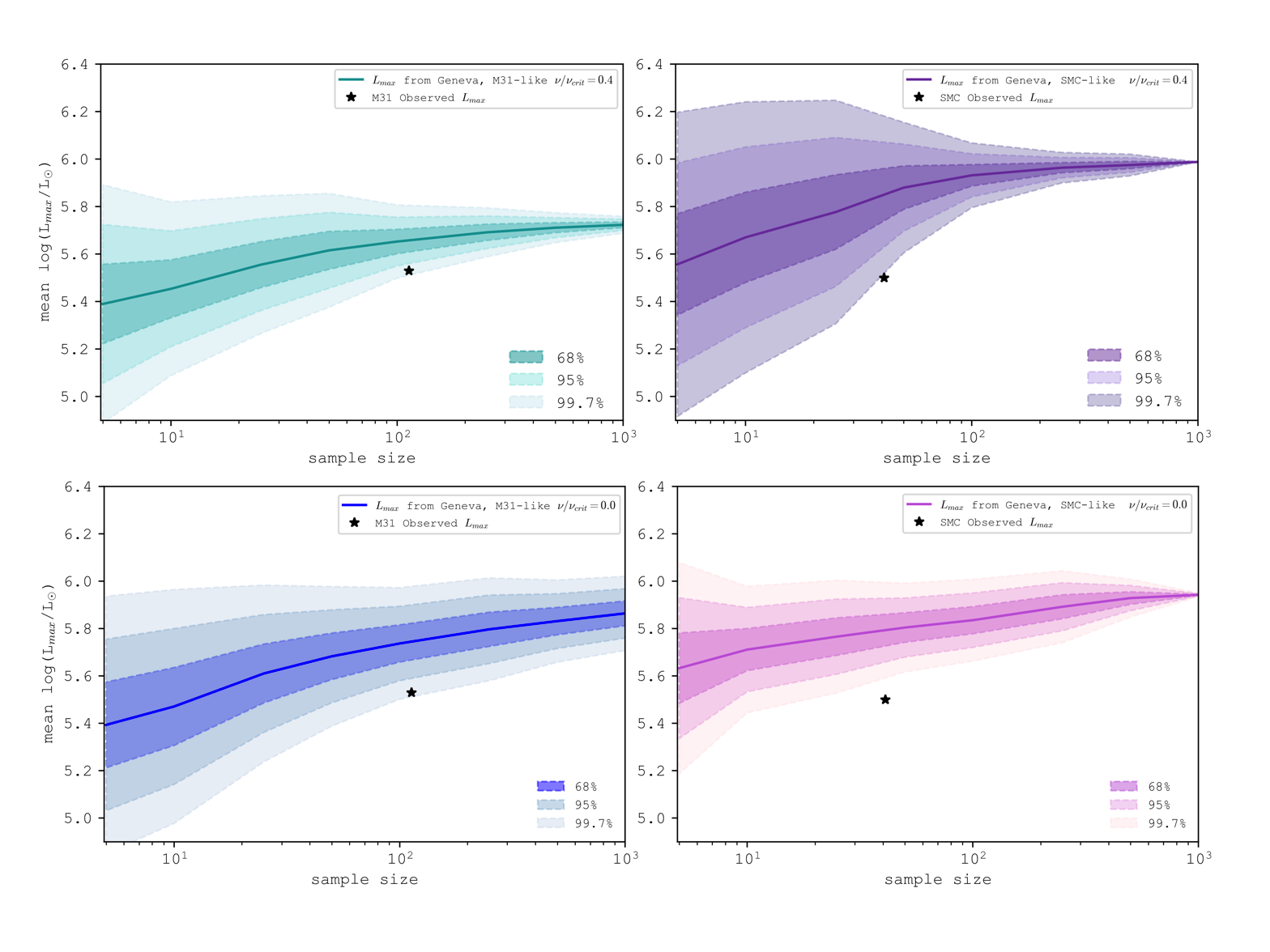}}
    \caption{The expected $L_{\rm max}$ for a range of sample sizes as predicted by the Geneva rotating models for both solar (Z=0.014) and SMC-like (Z=0.002) metallicities. The shaded regions indicate the confidence limits on $L_{\rm max}$ as shown in the legend and the black stars indicate the observed $L_{\rm max}$ and sample size for M31 from this work and the same for the SMC from \citet{DCB18}.}
    \label{lmax_vs_n}
\end{figure*}

\subsection{Comparison to lower metallicities}
\label{Comparison_to_lower_metallicities}
To make a broader test of the metallicity dependence of $L_{\rm max}$ and the luminosity function, we perform two comparisons. Firstly, we compare the empirical luminosity functions of the LMC and SMC with M31. Secondly, we compare the M31 luminosity function and $L_{\rm max}$ to theoretical expectations of lower metallicities using population synthesis. \newline

\subsubsection{Observational comparisons between the LMC and SMC}
\label{local_group_comparison}
We look at the cumulative RSG luminosity function for M31 and compare with the empirical SMC and LMC distributions from DCB18, looking at all RSGs with $\log(L/L_{\odot}) $>$ 5$, where our sample is considered to be complete. In these galaxies, the metallicities are thought to be $\sim 0.25Z\odot$ and $0.5Z\odot$, respectively \citep{russell&dopita90}. As noted previously, we assume that the M31 metallicity lies in the range of $1.06-1.66Z\odot$ \citep{zurita&bresolin12}. The left panel of Figure~\ref{2_panel_ks_test} shows the similarities of the observed cumulative luminosity functions for M31, SMC and LMC. We perform a Kolmogorov-Smirnov (KS) test to evaluate these similarities by measuring the differences between the cumulative distribution functions. We find for the empirical M31 distribution compared with the SMC and LMC, a 60\% and 44\% probability, respectively, that they are drawn from the same parent distribution. Hence, the probability that the RSG luminosity function in the three galaxies are consistent with one another is within $1\sigma$. Furthermore, each galaxy has the same $L_{\rm max}$ to within 0.1dex, at $\log(L/L_{\odot}) \sim 5.5$. Therefore, we find no evidence that the luminosities of RSGs have a dependence on metallicity. In the next section, we will compare these empirical findings to theoretical predictions.   \newline

\subsubsection{Theoretical predictions of the luminosity distribution}
\label{evo_models}
 To compare our observational results to theoretical predictions, we perform a population synthesis analysis. We do so by first generating a sample of random initial masses between $8-60M_{\odot}$ according to the Salpeter initial mass function (IMF). Each star is randomly assigned an age between 0 and 38 Myr, under the implicit assumption of a constant star-formation rate. We then match these to evolutionary tracks using the SYnthetic CLusters Isochrones \& Stellar Tracks (\textsc{syclist}) from the Geneva group at solar metallicity (Z=0.014)  \citep{ekstrom12}, to interpolate $L_{\rm bol}$ and $T_{\rm eff}$ from the track of each simulated star, removing any stars with age greater than the stars maximum expected lifetime. We also apply a temperature cut at log $T_{\rm eff} < 3.8$ to ensure the sample consists of cool supergiants. We perform a Monte Carlo experiment where we draw a random sample of stars from the model population, matching the observed number of RSGs in M31 and show the mean number of stars in each luminosity bin for both the rotating and non rotating models. The result is a simulated luminosity distribution for a constant star formation rate. 
 
 The comparison of this simulated distribution to the observations shows that the model predictions perhaps slightly over-predict the number of luminous stars at the high end of the distribution compared to observation for M31, but more notably, predicts $L_{\rm max}$ to be much higher than we observe. However, at the high luminosity end we do have a very small sample size and so our results are subject to stochastic uncertainties, which we will quantify in Section~\ref{Comparisons_to_theoretical_predictions_of_lmax}. \newline

The right panel of Figure~\ref{2_panel_ks_test} shows the model cumulative RSG luminosity functions at M31-like metallicity for both the rotating and non-rotating models. It can be seen that although the shape of the M31 cumulative distributions are quite similar, there is a distinct difference in $L_{\rm max}$, where the non-rotating models predict a higher maximum luminosity compared to the rotating models. When comparing to the observed M31 cumulative distribution in the left panel of Figure~\ref{2_panel_ks_test}, there is a clear difference between the model and observed distributions. \newline

When we take a look at the model cumulative luminosity function of RSGs at SMC-like (Z=0.002) metallicity (SMC-like tracks are from \citealt{Georgy2013}), seen in the right panel of Figure~\ref{2_panel_ks_test}, there is not only a clear difference between the distributions of the rotating and non-rotating models, but a distinct contrast between the model M31 and model SMC distributions. Therefore, the models predict that we should see a difference between the RSG luminosity functions of M31 and the SMC. However, despite the contrast in metallicity, the observed RSG cumulative distributions are consistent with each other to within $1\sigma$, as shown previously in Section~\ref{local_group_comparison}. \newline

We now compare the observational and model predicted M31 and SMC-like cumulative distributions using a KS test, as in the previous section. Here we find a probability of 5\% (rotating) and 0.1\% (non-rotating) for the M31 models compared with observations and a $0.02\%$ (rotating) and $10^{-6}\%$ (non-rotating) probability for the SMC models compared with observations. These low probabilities lead us to conclude that there is little similarity between the model distributions in the two galaxies and they are unlikely to be drawn from the same parent distribution. This is in sharp contrast with what we see in the empirical distributions of M31 and the SMC, which are statistically indistinguishable.  \newline

\subsubsection{Comparisons to theoretical predictions of $L_{\rm max}$}
\label{Comparisons_to_theoretical_predictions_of_lmax}
From our observational study of the M31 RSG population, as previously discussed, after the marginal candidate J004520.67+414717.3 with log$(L/L_{\odot})=5.75\pm0.11$, the next five most luminous stars span the range of $5.43 < \log(L/L_{\odot}) < 5.53$, suggesting an upper luminosity limit for M31 of log$(L/L_{\odot}) \approx 5.5$. In this section, we take a closer look at the statistical significance of $L_{\rm max}$ at M31 and SMC-like metallicities as predicted from the Geneva models.\newline

By simply looking at the parameter space occupied by the evolutionary tracks on a HR diagram, the Geneva models predict that $L_{\rm max}$ for M31 should be in the range $5.7 \lesssim \log(L/L_{\odot}) \lesssim 5.8$, yet we observe a much lower limit of $\approx 5.5$. However, we are dealing with small number statistics at the high luminosity end. This results in stochastic effects where the $L_{\rm max}$ we observe is a function of our sample size, meaning the larger the sample size, the higher the probability of sampling close to the true HD limit. Therefore, when comparing model predictions to observations, we must be careful to take this effect into account. \newline

To investigate the effects of sample size on $L_{\rm max}$, we perform another Monte Carlo experiment where we randomly select $N$ stars from the theoretical luminosity function and determine $L_{\rm max}$ of that sample. We repeat this $10^5$ times to find the average $L_{\rm max}$ for each sample size of $N$ cool supergiants with log$(L/L_{\odot})$ > 5. The results of this Monte Carlo are shown in Figure~\ref{lmax_vs_n} for both M31 and SMC-like metallicities. It shows the $L_{\rm max}$ we would expect to measure, plus the confidence intervals of that value, as a function of sample size. As one would expect, larger sample sizes result in the higher luminosity bins being more populated, meaning that the $L_{\rm max}$ we observe is more likely to reflect the `true' $L_{\rm max}$, with a smaller associated uncertainty. \newline

In each panel of Figure~\ref{lmax_vs_n}, the empirical $L_{\rm max}$ for the sample size we observe for that galaxy is denoted by the black star. Although M31 shows agreement within $3\sigma$, the SMC shows a disagreement beyond the 99.7\% confidence limit. This increasing disagreement between observations and theoretical predictions as a function of metallicity can be understood as follows: As shown earlier, the empirical $L_{\rm max}$ is observed to be metallicity-invariant. By contrast, the theoretical expectation of $L_{\rm max}$ in single star evolution is governed by metallicity-dependent mass-loss, and so increases with decreasing Z. \newline

In summary, we find no significant difference in $L_{\rm max}$ within the errors across a metallicity baseline of ($0.25Z_{\odot}$ to $\gtrsim Z_{\odot}$). This is in clear disagreement with theoretical expectations because $L_{\rm max}$ predictions from the models are simply too high compared to observational measurements and this effect is predicted to only increase with decreasing metallicity. \newline

\subsection{Possible explanations for a metallicity invariant HD limit}
The results of this work have shown that the observational luminosity function of RSGs do not follow theoretical expectations, both in terms of $L_{\rm max}$ and the shape of the luminosity function. There are several well-known sources of uncertainty in stellar evolutionary models, particularly in the pre-supernova phases of massive stars, such as mass-loss, mixing processes and rotational effects. In the present section, we discuss the possible implications these parameters may have for the theoretical predictions of the HD limit. \newline

\subsubsection{Mass loss}
\label{massloss}
Mass loss is a key process responsible for the stripping of the Hydrogen envelopes of stars. Hot star winds on the MS are driven by radiation pressure due to metal absorption lines in the UV, which means wind strength is sensitive to metallicity. It is this dependence of wind strength on metallicity which results in the predicted metallicity dependence of the HD limit in single star models. However, it has been seen from the cumulative luminosity functions of the RSGs in M31, LMC and SMC and from the invariance of $L_{\rm max}$ across these galaxies that there is no metallicity dependence. Also, recent work has shown that the mass loss rates from these metallicity dependent hot star winds are being revised downward by a factor of $\sim3$ \citep[e.g.][]{sundqvist2019,Bjorklund2021}, and so they are even less effective at removing the Hydrogen envelope than previously thought. Therefore, we conclude that line-driven winds in the hot star phases can \textbf{not} be the cause of the HD limit.  \newline

We next take a look at the contribution of mass loss as a result of RSG winds, for which there is some evidence to suggest that more metal poor environments result in weakened RSG wind speeds \citep[e.g.][]{goldman17}. The most widely used RSG wind prescription in stellar evolutionary codes is from \citet{Dejager88}, but is thought to over-estimate the rate of mass loss ($\dot{M}$), particularly for more luminous RSGs, as discussed in \citet{beasor2020}. A new RSG $\dot{M}$ prescription, presented in the latter study, implies that only a small fraction of envelope mass is lost during the RSG phase ($\sim1M_{\odot}$). This is considerably lower than with the prescription implemented in the Geneva models, in which up to $\sim$ 50\% of the envelope mass can be lost during this period. In fact, with the \citet{beasor2020} mass loss recipe implemented instead, higher mass stars ($> 30 M_{\odot}$) no longer evolve back to the bluer side of the HR diagram, resulting in a larger number of higher mass stars remaining in the RSG phase. Therefore, despite offering a more accurate description of $\dot{M}$ for cool supergiants in stellar models, in regards to the HD limit the disagreement actually worsens, giving rise to an even greater upper limit of $\log(L/L_{\odot}) \sim 6$. This means that RSG winds are simply not strong enough to be responsible for the HD limit.\newline

The lack of metallicity dependence means that line-driven winds cannot be responsible for the HD limit. However, this doesn't rule out the episodic type mass loss seen in Luminous Blue Variables (LBVs). LBV eruptive mass loss is so strong that the winds become optically thick and are likely to be driven by continuum radiation pressure in super-Eddington phases \citep{nathansmith2006}. Since we observe LBV eruptions at high and low metallicity, LBV mass loss is not metallicity dependent \citep{smith&owocki06}. This means we cannot rule out mass loss from LBV type eruptions as a potential cause of the HD limit. Similarly, \citet{Kraus2015} suggest that stars in the B[e] supergiant phase, are also thought to eject large amounts of material, much like LBVs which could be another possible type of mass loss contributing to the HD limit.

Further, it has been argued that the origin of LBV-type eruptions could be a consequence of binary interaction and mergers \citep[e.g.][]{nathansmith14}, which could also be an explanation for the existence of the HD limit (see next section).

\begin{figure}
 	\centerline{\includegraphics[width=1.12\columnwidth]{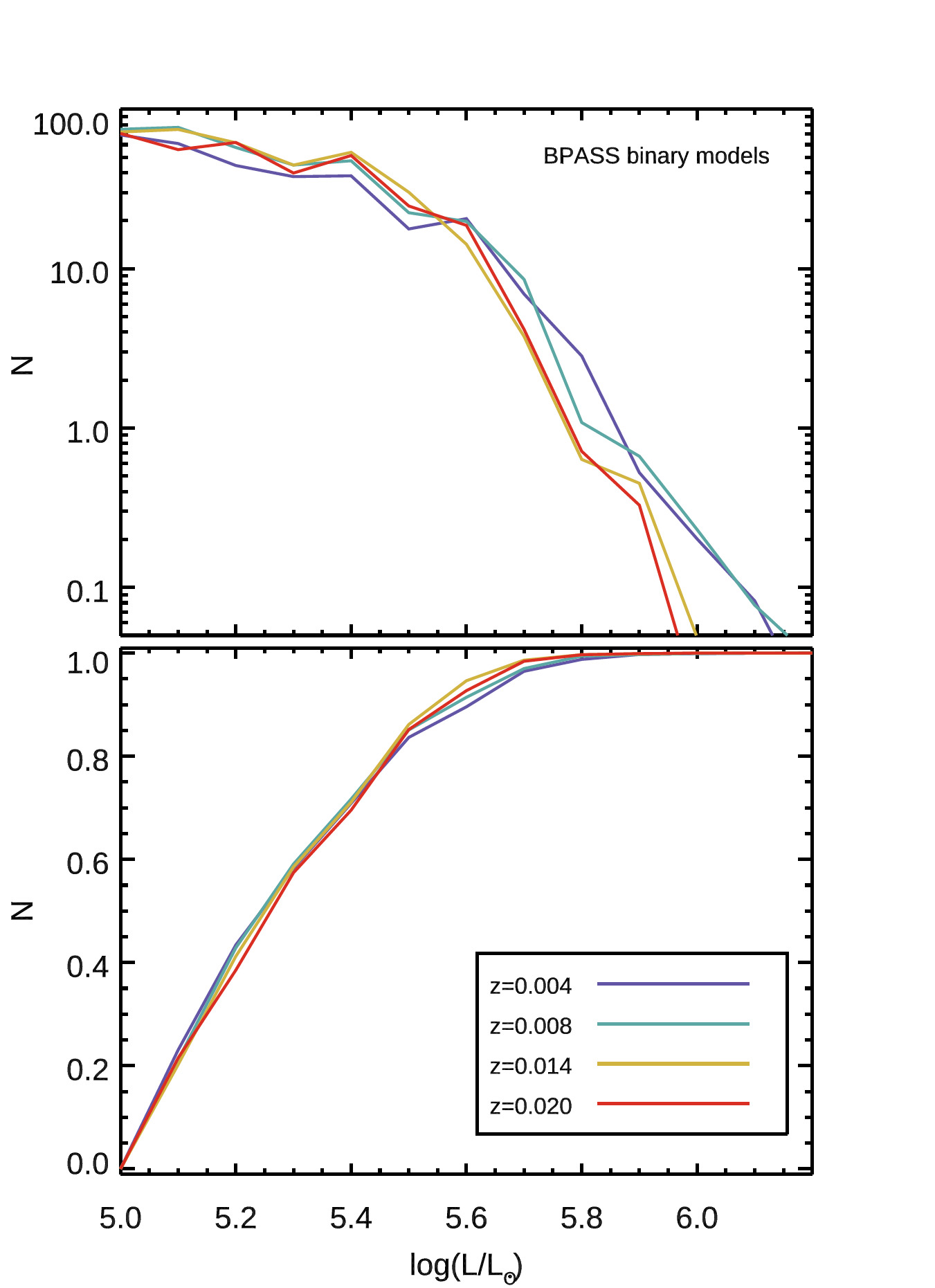}}
     \caption{Top panel: Predictions of the luminosity function of cool supergiants from BPASS binary population synthesis for the metallicity range Z=0.004 to Z=0.020. Bottom panel: The same result but shown as a cumulative luminosity distribution.}
     \label{bpass}
\end{figure}

\subsubsection{Binarity}
Thus far, in seeking to understand the RSG populations across the three galaxies, we have exclusively considered single-star evolutionary models. However, it is becoming increasingly clear that such models are of limited relevance for the most massive stars. Several studies in the literature have concluded that the fraction of OB stars in binary systems are in the range of 50-60\% or higher \citep{SANA2012,SANA2013,dunstall2015}. Furthermore, the probability of a star being in a multiple, \textit{and} that the star will interact with this companion, appears to increase with increasing mass \citep{Duch2013,moedistefano2017}. This is also suggested in the recent work of \citet{Bodensteiner2021}, who find that the bias-corrected close binary fraction of the $\sim40$ Myr old massive SMC open cluster, NGC330, is $34^{+8}_{-7}$\%. This is a lower fraction compared to younger clusters in the Milky Way and LMC. For example, the Cygnus OB2 association within our Galaxy has an intrinsic binary fraction of $\approx55$\% \citep{kobulnicky14}. The counterpart fraction for the overall B-star population of the LMC 30 Doradus region is found to be $58\pm11$\% \citep{dunstall2015}. This means that above some mass threshold, it is reasonable to expect that the likelihood of a star evolving according to single stellar evolutionary tracks will eventually tend towards zero. Specifically, a star's evolution to the red will be prevented by interaction with a companion either in or before the Hertzsprung Gap. This means we would expect binary effects to also be contributing to the mass lost during a star’s life and is therefore a possible explanation for the reduced $L_{\rm max}$ we see in observations \citep{DCB18}. \newline

To investigate the effects of binarity on $L_{\rm max}$, we extracted the RSG luminosity function for a constant star formation history from the Binary Population and Spectral Synthesis (BPASS) models. These models assume the mass ratio and period distributions from \citet{moedistefano2017}, which specify that stars with masses relevant for RSGs (M>9Msun) have close binary frequencies in excess of 80\%. The BPASS RSG luminosity functions as a function of Z are shown in Figure~\ref{bpass}. We still see a metallicity dependence and very high $L_{\rm max}$, similar to single star models, in the metallicity range Z=0.004 to Z=0.020.

Given the very high close binary fraction for massive stars set within the BPASS simulations, one would expect that most, if not all, of these stars would interact prior to the primary reaching the RSG phase. It is therefore intriguing that the BPASS-simulated RSG luminosity functions behave so similarly to those of the single star evolution models. In the future, it would be of interest to further mine the BPASS results to investigate the histories of the RSGs in these simulations. \newline

\section{Conclusions}
\label{conclusions}
We have compiled a sample of mid-IR selected cool supergiants to measure the luminosity function of the red supergiant (RSG) population in M31 to investigate the Humphreys-Davidson limit ($L_{\rm max}$).
\begin{itemize}
    \item We find that the luminosity function of RSGs is independent of metallicity, based on the range of metallicities studied here (from SMC-like to M31-like).  \newline
    \item $L_{\rm max}$ is also independent of metallicity, where we find the HD limit for M31 is $\log(L/L_{\odot}) =5.53 \pm 0.03$, within 0.1dex of the SMC and LMC. We are in agreement with \citet{DCB18} who find a lack of evidence for a metallicity dependent $L_{\rm max}$. This suggests that mass loss from line-driven winds are \textbf{not} the cause of the HD limit. \newline
    \item A population synthesis analysis shows that the single star Geneva evolutionary models not only over-predict the number of luminous cool supergiants at the high luminosity end, but also over-predict $L_{\rm max}$, particularly at lower metallicities.\newline

\end{itemize}

\section*{Acknowledgements}
The authors would like to thank the referee for their useful comments. SLEM thanks Nathan Smith, Gemma González-Torà and Tom Sedgwick for useful comments and insightful conversations. SLEM acknowledges support from an STFC studentship, jointly supported by the Faculty of Engineering and Technology at the Astrophysics Research Institute at Liverpool John Moores University. ERB is supported by NASA through Hubble Fellowship grant HST-HF2-51428 awarded by the Space Telescope Science Institute, which is operated by the Association of Universities for Research in Astronomy, Inc., for NASA, under contract NAS5-26555. This research has made use of the IDL software/astrolib,  Matplotlib \citep{matplotlib}, NumPy \citep{numpy}, SciPy \citep{2SciPy}, as well as the VizieR catalogue access tool and the SIMBAD/Aladin Sky Atlas operated at CDS, Strasbourg, France \citep{vizier,simbad, aladin1,aladin2}. 

\section{Data Availability}
The data underlying this article will be shared on reasonable request to the corresponding author. The datasets were derived from sources in the public domain: 
\begin{itemize}
    \item {LGGS: \\
    \href{{https://cdsarc.unistra.fr/viz-bin/cat/J/AJ/152/62}} {\url{cdsarc.unistra.fr/viz-bin/cat/J/AJ/152/62}}} \newline
     \item {GaiaEDR3: \\
     \href{{https://cdsarc.unistra.fr/viz-bin/cat/I/350}} {\url{cdsarc.unistra.fr/viz-bin/cat/I/350}}} \newline
    \item {2MASS: \\
    \href{{https://cdsarc.unistra.fr/viz-bin/cat/II/246}} {\url{cdsarc.unistra.fr/viz-bin/cat/II/246}}} \newline
    \item {Spitzer M31 point source survey:\\ \href{https://cdsarc.unistra.fr/viz-bin/cat/J/ApJS/228/5}{\url{cdsarc.unistra.fr/viz-bin/cat/J/ApJS/228/5}}} \newline
\end{itemize}

\bibliography{main.bib}

\appendix

\label{lastpage}
\end{document}